\documentclass[apj]{emulateapj}

\usepackage{longtable}
\usepackage{multirow}
\usepackage{color}

\shorttitle{Models of kHz QPOs vs. NS EoS}
\shortauthors{G. T\"{o}r\"{o}k et al.}

\def\sss{\scriptscriptstyle}
\def\U{{\sss \!U}}
\def\L{{\sss \!L}}
\def\K{{\sss \!K}}
\def\LT{{\sss \!L\,\!T}}

\def\nur{\nu_\mathrm{r}}
\def\nuv{\nu_\theta}
\def\nuL{\nu_\L}
\def\nuU{\nu_\U}
\def\nuK{\nu_\K}
\def\nuLT{\nu_\LT}

\def\B{\mathcal{F}}

\begin{document}


\title{Constraining models of twin peak quasi-periodic oscillations \\ with realistic neutron star equations of state}


\author{Gabriel T\"{o}r\"{o}k$^1$, Kate\v{r}ina Goluchov\'{a}$^{1,2}$, Martin Urbanec$^1$, Eva \v{S}r\'{a}mkov\'{a}$^1$, \\ 
			Karel Ad\'{a}mek$^{1,2,3}$, Gabriela Urbancov\'{a}$^1$, Tom\'{a}\v{s} Pech\'{a}\v{c}ek$^1$, Pavel Bakala$^1$, Zden\v{e}k Stuchl\'{\i}k$^2$, Ji\v{r}\'{\i} Hor\'ak$^4$, Jakub Jury\v{s}ek$^{1,5}$
        }
				
\affil{$^1$ Research Centre for Computational Physics and Data Processing, Institute of Physics, Faculty of Philosophy \& Science, Silesian University in Opava, Bezru\v{c}ovo n\'am.~13, CZ-746\,01 Opava, Czech Republic \\
$^2$ Research Centre for Theoretical Physics and Astrophysics, Institute of Physics, Faculty of Philosophy \& Science, Silesian University in Opava, Bezru\v{c}ovo n\'am.~13, CZ-746\,01 Opava, Czech Republic  \\
$^3$ University of Oxford, Oxford e-Research Centre, 7 Keble Road, Oxford, OX1 3QG \\
$^4$ Astronomical Institute, Bocni II 1401/2a, CZ-14131 Praha 4 - Sporilov, Czech Republic \\
$^5$ Astronomical Institute, Charles University Prague, Faculty of Mathematics and Physics,
V Hole\v{s}ovi\v{c}k\'{a}ch 2, Praha 8, CZ-180 00, Czech Republic 
}


\email{Mailto: gabriel.torok@gmail.com}



\begin{abstract}
Twin-peak quasi-periodic oscillations (QPOs) are observed in the X-ray power-density spectra of several accreting low-mass neutron star (NS) binaries. In our previous work we have considered several QPO models. We have identified and explored {mass--angular-momentum relations} implied by individual QPO models  for the atoll source 4U~1636-53. In this paper we extend our study and confront QPO models with various NS equations of state (EoS). We start with simplified calculations assuming Kerr background geometry and then present results of detailed calculations considering the influence of NS quadrupole moment (related to rotationally induced NS oblateness) assuming Hartle-Thorne spacetimes. We show that the application of concrete EoS together with a particular QPO model yields a specific mass--angular-momentum relation. However, we demonstrate that the degeneracy in mass and angular momentum can be removed when the NS spin frequency inferred from the X-ray burst observations is considered. {We inspect a large set of EoS and discuss their compatibility with the considered QPO models.} We conclude that when the NS spin frequency in 4U~1636-53 is close to 580Hz we can exclude {51} from {90} of the considered combinations of EoS and QPO models. We also discuss additional restrictions that may exclude even more combinations. Namely, there are {13} EOS compatible with the observed twin peak QPOs and the relativistic precession model. However, when considering the low frequency QPOs and Lense-Thirring precession, only {5} EOS are compatible with the model.
\end{abstract}





\section{Introduction}\label{intro}

Accreting neutron stars (NS) are believed to be the compact component in more than 20 low mass X-ray binaries (LMXBs). In these systems, the mass is transferred from the companion by overflowing the Roche lobe and forming an accretion disc that surrounds the NS. The disc contributes significantly to high X-ray luminosity of these objects while most of the radiation comes from its inner parts and the disc--NS boundary layer. According to their X-ray spectral and timing properties, the NS LMXBs have been further classified into Z and atoll sources, whose names have been inspired by the shapes of tracks they trace in the color-color diagram \citep[e.g.][]{kli:2005}. While the Z sources are generally more stable and brighter, the atoll sources are weaker and show significant changes in the X-ray luminosity. Both classes exhibit a variability over a large range of frequencies. Apart from irregular changes, their power spectra contain also relatively coherent features known as quasi-periodic oscillations (QPOs).

The so-called low frequency QPOs have frequencies in the range of $1-100$Hz. In the case of Z-sources they have been further classified into horizontal, flaring, and normal branch oscillations (HBO, FBO and NBO, respectively) depending on position of the source in the color-color diagram. Oscillations of properties similar to HBOs have been observed also in several atoll sources \citep[see][ for a review]{kli:2006}. Much attention among theoreticians is however attracted to the kilohertz QPOs ($100-1000$Hz) because their high frequencies are comparable to orbital timescale in the vicinity of a NS. It is believed that this coincidence represents a strong indication that the corresponding signal originates in the innermost parts of the accretion discs or close to the surface of the NS itself. This belief has been supported also by the means of the Fourier-resolved spectroscopy \citep[e.g.,][]{gilf-etal:2000}.

The kHz QPOs have similar properties in both Z and atoll sources. They are frequently observed in pairs often called twin peak QPOs. Their `upper' and `lower' QPO frequencies ($\nuU$ and $\nuL$, respectively) exhibit a strong and remarkably stable positive correlation and clustering around the rational ratios. These ratios are emphasized either due to the intrinsic source clustering, or due to weakness of the two QPOs outside the limited frequency range \citep[suggesting possible resonant energy exchange between two physical oscillators,][]{abr-etal:2003a, bel-etal:2005,bel-etal:2007,  tor-etal:2008a, tor-etal:2008b, tor-etal:2008c, bar-bou:2008, hor-etal:2009, bou-etal:2010}. Other properties of each oscillation (e.g. the rms-amplitude and the quality factor) seem to mostly depend on its frequency, and the way how they vary is different between the upper and lower oscillation. These differences often help to identify the type of kHz QPO in cases when only one peak is present in the power spectra \citep{bar-etal:2005,bar-etal:2006,men:2006,tor:2009}. 

Many models have been proposed to explain the rich phenomenology of twin peak QPOs \citep[][ and several others]{alp-sha:1985,lam-etal:1985,mil-etal:1998a,psa-etal:1999b,wag:1999,wag-etal:2001,abr-klu:2001,klu-abr:2001,kat:2001,tit-ken:2002,abr-etal:2003b,abr-etal:2003c,rez-etal:2003,klu-etal:2004,pet:2005a,zha:2005,bur:2005,tor-etal:2007,kat:2007,kat:2008,stu-etal:2008,cad-etal:2008, kos-etal:2009, ger-etal:2009,muk:2009,stu-etal:2013,stu-etal:2014,torok-etal:2015:MNRAS:,wan-etal:2015:MNRAS:,stu-etal:2015}. While any acceptable model should address both the excitation mechanism and subsequent modulation of the resulting X-ray signal as well as their overall observational properties, most of the theoretical effort has been so far devoted to the observed frequencies. Clearly, their correlations serve as a first test of the model viability.

\subsection{Aims and Scope of this Paper}

Comparison between the observed and the expected frequencies can reveal the mass and angular momentum of the NS. These can be confronted with models of rotating NS based on a modern equation of state \citep[EoS, e.g.][]{urb-etal:2010:b}. In \cite{tor-etal:2012} we have identified and explored {mass--angular-momentum relations} implied in Kerr spacetimes by individual QPO models. We have also discussed that the degeneracy in mass and angular momentum can be removed when the NS spin frequency is known.


Here we extend our study and confront QPO models with a large set of NS equations of state (EoS) while focusing on the influence of NS quadrupole moment related to its rotationally induced oblateness. The paper is arranged as follows. In Section~\ref{section:fits:Kerr} we very shortly remind individual QPO models considered in our study and recall previously obtained results. We present here the completed simplified calculations that assume Kerr background geometry and the atoll source 4U~1636-53. These follow previous comparison between predictions of the relativistic precession model and 5 EoS. The consideration is extended to other models and a large set of 18 EoS. Sections~\ref{section:HTRP} and ~\ref{section:EoS} bring detailed consequent calculations of RP model predictions considering the influence of NS quadrupole moment within Hartle-Thorne spacetimes. We show here that the application of concrete Sly-4 EoS within the model in Hartle-Thorne spacetime  brings a specific mass--spin relation. This relation is confronted with the NS spin frequency inferred from the X-ray burst observations. In Section~\ref{section:conclusions} we present analogical results for the whole set of 5 QPO models and 18 EoS and outline their implications. We also discuss here the implications of the consideration of low frequency QPOs.

\begin{figure*}[t]
\begin{center}
\includegraphics[width=.9\linewidth]{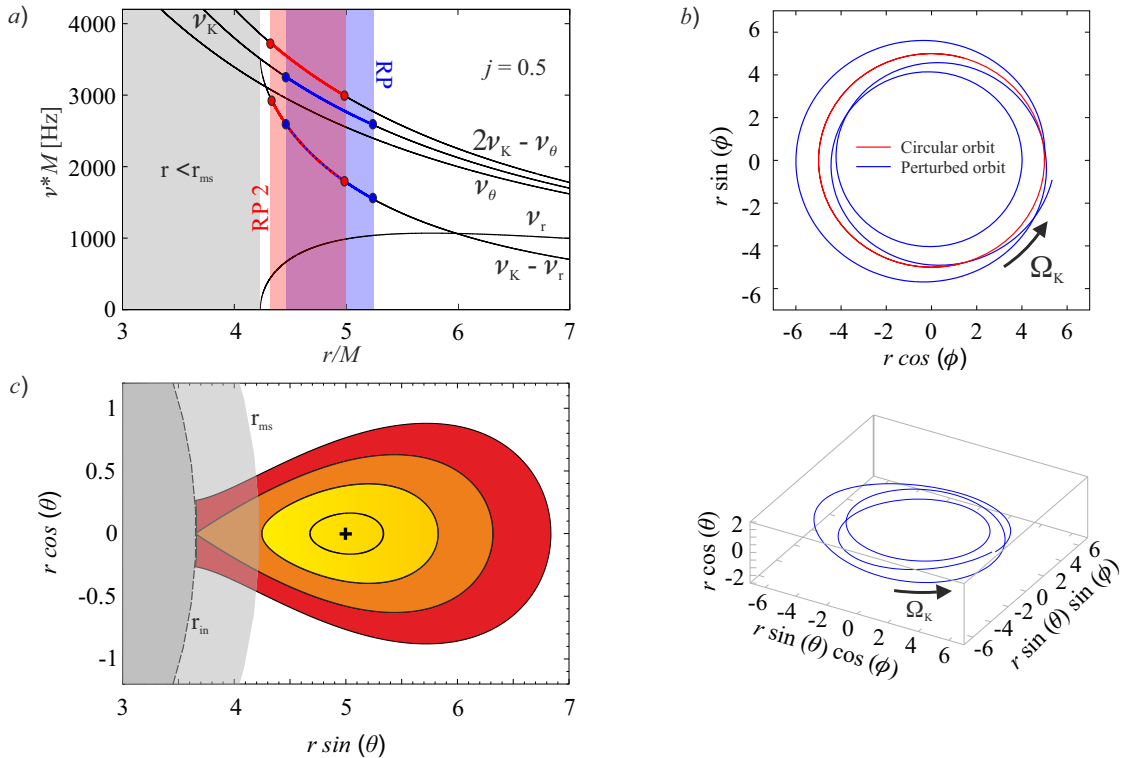}
\end{center}
\caption{Frequencies of orbital motion and illustration of models of QPOs in the atoll source 4U-1636-53. a) Behaviour of characteristic orbital frequencies in Kerr spacetimes. The blue area denotes a radial region associated to the RP model, i.e., the region where orbital and periastron precession frequencies can be identified with the frequencies observed in the atoll source 4U-1636-53.  The red area denotes the same but for the RP2 model frequencies. The grey area corresponds to the region below marginally stable circular orbit, $r<r_{\mathrm{ms}}$. b) Example of free test particle trajectory and its  projection onto the equatorial plane. Figure corresponds to the situation drawn in panel a) and RP model ($r = 5 \, M$). The red circle indicates an unperturbed circular trajectory.  c) Equipotencial surfaces determining the shape of torus located at $r = 5 \, M$ drawn for different values of torus thickness $\beta$. Slender torus limit ($\beta=0$) is denoted by the black cross. In this limit, and when the RP2 model is assumed, the torus oscillates with frequencies $\nuK(r) - \nur(r)$ and $2\nuK(r) - \nuv(r)$. In the limit of $j=0$ these frequencies coincide with the RP model frequencies $\nuK(r) - \nur(r)$ and $\nuK(r)$. Although the two models predict the same frequencies in the limit of non-rotating NS, the associated physical mechanisms are not the same.}
\label{figure:1}
\end{figure*}

\section{Twin Peak QPO Models Approximated in Kerr Spacetimes} 
\label{section:fits:Kerr}

Within the framework of many QPO models, the observable frequencies can be expressed directly in terms of epicyclic frequencies. Formulae for the geodesic Keplerian, radial and vertical epicyclic frequencies in Kerr spacetimes were first derived by \cite{ali-gal:1981}. In a commonly used form \citep[e.g.,][]{tor-stu:2005} they read
\begin{equation}\label{eq.orb}
\Omega_\K=\frac{\B}{j+{x}^{3/2}},\quad\nur=\Gamma\Omega_\K,\quad\nuv=\Delta\Omega_\K\,,
\end{equation}
where
\begin{eqnarray}
\Gamma&=&\sqrt{\frac{-3 j^2+8 j \sqrt{{x}}+\left(-6+{x}\right) {x}}{{x}^2}}, \\
\Delta&=&\sqrt{1+\frac{j \left(3 j-4 \sqrt{x}\right)}{x^2}}\,,
\end{eqnarray}
$x\equiv r/M$, and the "relativistic factor" $\B$ reads $\B \equiv c^3/(2\pi GM)$. We note that Kerr geometry represents an applicable approximation of NS spacetimes when the compact object mass is high \citep{tor-etal:2010,urb-etal:2013}.


The above formulae valid for Kerr spacetimes well describe (epicyclic) slightly perturbed circular geodesic motion. This description of epicyclic motion of test particles relevant to standard thin accretion discs may well approximate also epicyclic motion in fluid accretion flow provided that the pressure effects in the fluid are negligible and linear quasi-incompressible modes are considered. Formulae for geodesic epicyclic oscillations are often assumed within several QPO models based on accretion disc hot-spot as well as global fluid motion \citep[e.g.,][]{Ste-Vi:1999,ste-etal:2001,abr-klu:2001,klu-abr:2002}. Here we investigate a subset of models that have been previously considered in the study of \cite{tor-etal:2012}. 


\begin{table*}[ht]

\caption{ Models examined in this work.   \label{table:0}}

\begin{center}
\renewcommand{\arraystretch}{1.8}
{\begin{tabular}{cllcc}\tableline \tableline
Model & Relations & \multicolumn{3}{c}{$\nuL$ -  $\nuU$ relation}  \\ \tableline 
\multirow{2}{*}{RP} & $\nuL =  \nuK - \nur$,  &\multicolumn{3}{c}{\multirow{2}{*}{$\nuL = \nuU\left\{1 - \left[1 + \frac{8j\nuU}{\B - j\nuU} - 6\left(\frac{\nuU}{\B - j\nuU}\right)^{2/3} - 3j^2\left(\frac{\nuU}{\B - j\nuU}\right)^{4/3} \right]^{1/2}\right\}$}} \\
&$\nuU = \nuK$ &  && \\ \tableline
\multirow{2}{*}{TD} & $\nuL = \nuK$, & \multicolumn{3}{l}{\multirow{2}{*}{$\nuU = \nuL\left\{1 + \left[1 + \frac{8j\nuL}{\B - j\nuL} - 6\left(\frac{\nuL}{\B - j\nuL}\right)^{2/3} - 3j^2\left(\frac{\nuL}{\B - j\nuL}
\right)^{4/3} \right]^{1/2}\right\}$}}\\
& $\nuU = \nuK + \nur$ & &&\\ \tableline

WD & $\nuL = 2(\nuK - \nur)$, & $\nuU = 2\nuK - \nur$ &&\\ 
RP1 & $\nuL = \nuK - \nur$, &$\nuU = \nuv$ &&\\
RP2 & $\nuL = \nuK - \nur$, & $\nuU = 2\nuK - \nuv$ &&\\ 
\tableline
\end{tabular}}

\end{center}

\end{table*}

\subsection{Individual Models of QPOs and their Predictions}

Relativistic precession ({RP}) model explains the kHz QPOs as a direct manifestation of modes of relativistic epicyclic motion of blobs at various radii $r$ in the inner parts of the accretion disc \citep{Ste-Vi:1999}. For the RP model, one can easily solve relations defining the upper and lower QPO frequencies in terms of the orbital frequencies arriving at explicit formula which relates the upper and lower QPO frequencies in the units of Hertz \citep{tor-etal:2010}. We show this relation in Table~\ref{table:0}.  The concept of Tidal Disruption (TD) model is similar to the RP model, but the QPOs are atributed to a disruption of large accreting inhomogenities \citep{ger-etal:2009}. Evaluation of the explicit relation between the two observed QPO frequencies can be done in a way similar to the case of the RP model \citep{tor-etal:2012}, and we also include this relation in Table~\ref{table:0}.

While the former two models assume motion of a hot-spot propagating within the accretion disc, the Warp Disc (WD) model assumes non-axisymmetric oscillation modes in a thick disc \citep{kat:2001}. The two more considered models, RP1 and RP2, also deal with non-axisymmetric disc-oscillation modes. Frequencies of these modes coincide with the frequencies predicted by the RP model in the limit of $j=0$ \citep{bur:2005,tor-etal:2010}. Although the relevant frequencies coincide in the case of non-rotating NS, they correspond to a different physical situation (see Figure~\ref{figure:1} for an illustration). We include the expressions for lower and upper QPO frequency for all the three disc oscillation models in Table~\ref{table:0}.

\begin{figure*}
\begin{center}
\includegraphics[width=0.8\linewidth]{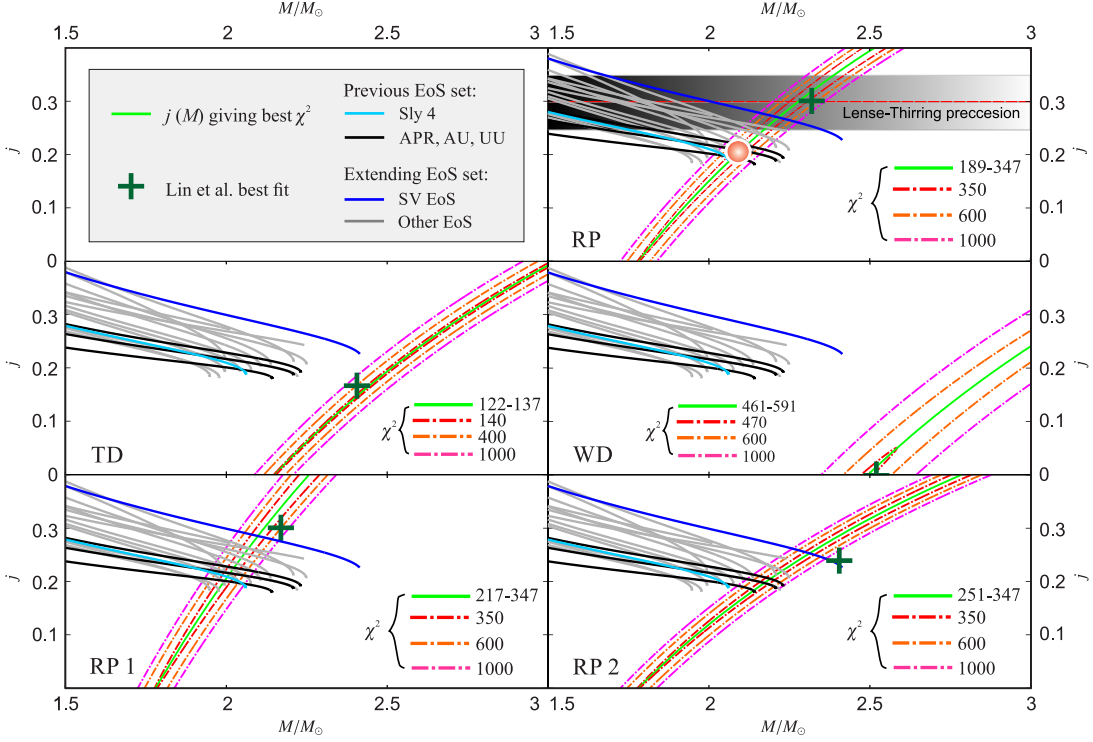}
\end{center}

\caption{$\chi^2$ maps {(20 d.o.f.)} calculated from data of the atoll source 4U~1636-53 and individual QPO models within Kerr spacetimes  vs. mass--angular-momentum relations predicted by NS EoS. For the calculations we consider 14 more EOS in addition to 4 EoS (SLy 4,  APR, AU-WFF1 and  UU-WFF2) assumed in \cite{tor-etal:2012}. The full set of 18 EoS is listed in Table~\ref{table:1}. In each panel the green line indicates the best $\chi^2$ for a fixed $M$ while the dashed green line denotes its quadratic approximation. The white lines indicate the corresponding 1$\sigma$ and 2$\sigma$ confidence levels.  The NS EoS are assumed for the rotational frequency of 580Hz inferred from the X-ray burst observations. The green cross-markers denote the mass and angular momentum
combinations reported for 4U 1636-53 and individual QPO models by \cite{Lin-etal:2011}.  The red spot roughly indicates the combination of mass and spin inferred from the common consideration of the RP model, NS spin frequency of $580$Hz and 4 EoS as discussed by \cite{tor-etal:2012}. Horizontal dashed red line together with the horizontal shaded bar indicate additional restrictions on the RP model following from consideration of Lense-Thirring precession as discussed in \cite{tor-etal:2012}.
}
\label{figure:2}
\end{figure*}

In  \cite{tor-etal:2010,tor-etal:2012} we assumed high mass (Kerr) approximation of NS spacetimes and relations from Table~\ref{table:0}. We have demonstrated that for each twin-peak QPO model and a given source the model consideration  results in a specific relation between the NS mass $M$ and angular momentum $j$ rather than in their single preferred combination. We pay a special attention to the atoll source 4U~1636-53 and evaluated mass--angular-momentum relations for all discussed QPO models.\footnote{\cite{Lin-etal:2011} have performed a similar analysis assuming a different set of twin-peak QPO frequency datapoints for the atoll source 4U-1636-53. The datapoints in their study have been obtained via common processing of a large amount of data while the datapoints used by \cite{tor-etal:2012} correspond to individual continuous observations of the source. It was shown in \cite{tor-etal:2012} that results of both studies are consistent (see also NS parameteres resulting within the two studies denoted in Figure \ref{figure:2} and the paper of \citeauthor{torok-etal:2015:MNRAS:}, \citeyear{torok-etal:2015:MNRAS:}).}  

\begin{table}\scriptsize

\caption{EoS examined in this work.   \label{table:1}}

\begin{center}

{\begin{tabular}{lllll}\tableline \tableline
\tabletypesize{\scriptsize}

 &   & $R$  & $n_\mathrm c$  & \\ 
EoS & $M_\mathrm{max}$ & (km) & ($[fm]^{-3}$) &  Reference  \\\tableline
Sly4    &  2.04 & 9.96 & 1.21 & 1 \\
SkI5   & 2.18 & 11.3& 0.97 & 1 \\
SV      & 2.38$^*$ & 11.9 & 0.80  & 1 \\
SkO    & 1.97 & 10.3 & 1.19 & 1  \\
Gs     & 2.08 & 10.8 & 1.07 & 1 \\
SkI2   & 2.11 & 11.0 & 1.03 & 1 \\
SGI    & 2.22 & 10.9 & 1.01 & 1 \\
APR    & 2.21 & 10.2 & 1.12 & 2 \\
AU     & 2.13 & 9.39 & 1.25 & 3 \\
UU     & 2.19 & 9.81 & 1.16 & 3 \\
UBS   &  2.20$^*$ & 12.1 & 0.68  & 4 \\ 
GLENDNH3 & 1.96 & 11.4 & 1.05  & 5 \\
Gandolfi & 2.20 & 9.82 & 1.16 & 6 \\
QMC700 & 1.95 & 12.6 & 0.61 & 7 \\
KDE0v1 & 1.96 & 9.72 & 1.29 & 8 \\
NRAPR & 1.93 & 9.85 & 1.29 & 9 \\
PNM L80 & 2.02 & 10.4 & 1.16 & 10 \\
J35 L80 & 2.05 & 10.5 & 1.14 & 10 \\ \tableline
\end{tabular}}
\tablecomments{ The individual columns indicate the maximum mass and the corresponding radius, and the central baryon number density for each EoS along with the relevant references. The asterisk symbols mark three EoS  whose maximum mass corresponds to maximum density allowed by the available EoS--table, and not to marginally stable star.}
\tablerefs{
(1) \cite{rik-etal:2003} (2) \cite{akm-etal:1998} (3) \cite{wir-etal:1988}  (4) \cite{urb-etal:2010:a} (5) \cite{glendnh3} (6) \cite{gandolfi} (7) \cite{qmc700} (8) \cite{kde0v1} (9) \cite{nrapr}
(10) \cite{will} 
}
\end{center}

\end{table}


{\subsection{Twin Peak QPO Models vs. NS EoS}}

In \cite{tor-etal:2012} we compared $\chi^2$ map describing the quality of the RP model fit of the 4U 1636-53 data to the $M-j$ relations implied by 5 specific NS EoS.  These $M-j$ relations have been calculated assuming that the NS spin frequency $\nu_{\mathrm{S}}$ is  580Hz \citep[][]{stro-Mar:2002,wat:2012:ARAA,gal-etal:2008:APJS:}. In those calculations we have utilized the approach of \cite{har:1967,har-tho:1968, cha-mil:1974, mil:1977,urb-etal:2010:a}.

In the top panel of Figure~\ref{figure:2} we show comparison between predictions of the RP model and 4 EoS carried out in \cite{tor-etal:2012}. {We note that the choice of  concrete EoS utilized within that paper was motivated by low values of a scaled quadrupole moment $\tilde{q}\equiv q/j^2$ of the assumed NS configurations.\footnote{In \cite{tor-etal:2012} we also assumed one more EOS {(WS, \cite{wir-etal:1988,ste-fri:1995})}. We do not consider this EOS here since it does not fulfill requirements of present observational tests.}} Although QPO model predictions are drawn for simplified calculations assuming Kerr background geometry, in following we do not restrict ourselves to high mass (compactness) NS. We thus add 14 more EoS which are indicated within the Figure. The full set of 18 EoS considered hereafter is listed in Table~\ref{table:1}. In the other panels of Figure~\ref{figure:2} we make the same comparison but for the other four considered QPO models.

In \cite{tor-etal:2012} we made a direct comparison between (just a few) EoS and the RP model. Inspecting our overall extended Figure~\ref{figure:2} we can expect that QPO models put strong restrictions on NS parameters and EoS, or vice versa. For instance, direct confrontation of EoS and TD model predictions strongly suggests that the model \cite[favoured within the study of][]{Lin-etal:2011} does not at all meet requirements given by the consideration of NS EoS. Moreover, comparing overlaps between the RP model relation and curves denoting the requirements of individual EoS, other interesting information can be obtained. Namely, there is a difference between overlaps considered in \cite{tor-etal:2012} denoted here by the red spot in Figure~\ref{figure:2} and overlaps given by the newly considered EoS. Clearly, the high quadrupole moment of NS configurations related to the latter set of EoS increases the required NS angular momentum. For instance, there is $j\doteq0.19$ for Sly4 vs. $j\doteq0.28$ for SV EoS. It is also apparent that such effect can be important for the consideration of Lense-Thirring precession and low frequency QPOs within the framework of RP model.

Motivated by these findings, in next we explore restrictions on QPO models in detail and perform consistent calculations in Hartle-Thorne spacetimes.

\bigskip

\section{Calculations in Hartle-Thorne Spacetimes}
\label{section:HT}

So far we have considered only Kerr approximation of the rotating NS spacetime assuming that the star is very compact. In such case the NS quadrupole moment $q$ related to its rotationally induced oblateness reaches low values and we have $\tilde{q} \approx 1$. In a more general case of $\tilde{q}>1$ one should assume NS spacetime approximated by the Hartle-Thorne geometry \citep{har:1967,har-tho:1968}.\footnote{The adopted approximation represents a convenient alternative to (more precise) numerical approach \citep[discussed in the same context by][]{ste-etal:1999} or other spacetime descriptions \citep[e.g.][]{Manko-etal:2000:,stu-gam:2002:,pap:2015}, see also \cite{bon-etal:1993,ste-fri:1995,bon-etal:1998,noz-etal:1998,ans-etal:2003,ber-etal:2005}.}

Based on the Hartle-Thorne approximation, the Keplerian orbital frequency can be expressed as \citep{abr-etal:2003a}
\begin{eqnarray}\label{eq:Kepler}
\Omega_\K=\frac{\mathcal{F}}{x^{3/2}}\left[1- \frac{j}{x^{3/2}}+j^2F_{1}(x)+qF_{2}(x)\right],
\end{eqnarray}
where
\begin{eqnarray}
F_{1}(x)&=&[48-80x+4x^2-18x^3+40x^4+10x^5 \nonumber \\ 
&& +15x^6-15x^7](16(x-2)x^4)^{-1}+A(x),\nonumber\\
F_{2}(x)&=&\frac{5(6-8x-2x^2-3x^3+3x^4)}{16(x-2)x}-A(x),\nonumber\\
A(x)&=&\frac{15(x^3-2)}{32}\ln\left(\frac{x}{x-2}\right).\nonumber
\end{eqnarray}
The radial and vertical epicyclic frequencies are then described by the following terms
\begin{eqnarray}
\nur^2&=&\frac{\mathcal{F}^{2}(x-6)}{x^{4}}[1+
jH_{1}(x)-j^2H_{2}(x)-qH_{3}(x)],
\label{eq:epicrad}\\
\nuv^2&=&\frac{\mathcal{F}^{2}}{x^{3}}[1- jG_{1}(x)+j^2G_{2}(x)+qG_{3}(x)],
\label{eq:epicvert}
\end{eqnarray}
where
\begin{eqnarray}
H_{1}(x)&=&\frac{6(x+2)}{x^{3/2}(x-6)},\nonumber\\
H_{2}(x)&=&[8x^4(x-2)(x-6)]^{-1}[384-720x-112x^2- \nonumber\\
&& 76x^3\nonumber\\
&&-138x^4-130x^5+635x^6-375x^7+60x^8]\nonumber\\
&& +C(x),\nonumber\\
H_{3}(x)&=&\frac{5(48+30x+26x^2-127x^3+75x^4-12x^5)}{8x(x-2)(x-6)}\nonumber\\
&& -C(x),\nonumber\\
C(x)&=&\frac{15x(x-2)(2+13x-4x^2)}{16(x-6)}\ln\left(\frac{x}{x-2}\right),\nonumber\\
G_{1}(x)&=&\frac{6}{x^{3/2}},\nonumber\\
G_{2}(x)&=&[8x^4(x-2)]^{-1}[48-224x+28x^2+6x^3-170x^4\nonumber\\
&&+295x^5 -165x^6+30x^7]-B(x),\nonumber\\
G_{3}(x)&=&\frac{5(6+34x-59x^2+33x^3-6x^4)}{8x(x-2)}+B(x),\nonumber\\
B(x)&=&\frac{15(2x-1)(x-2)^2}{16}\ln\left(\frac{x}{x-2}\right).\nonumber
\end{eqnarray}

\subsection{Results for the RP Model} \label{section:HTRP}

\begin{figure*}[t]
\begin{center}
\includegraphics[width=0.95\linewidth]{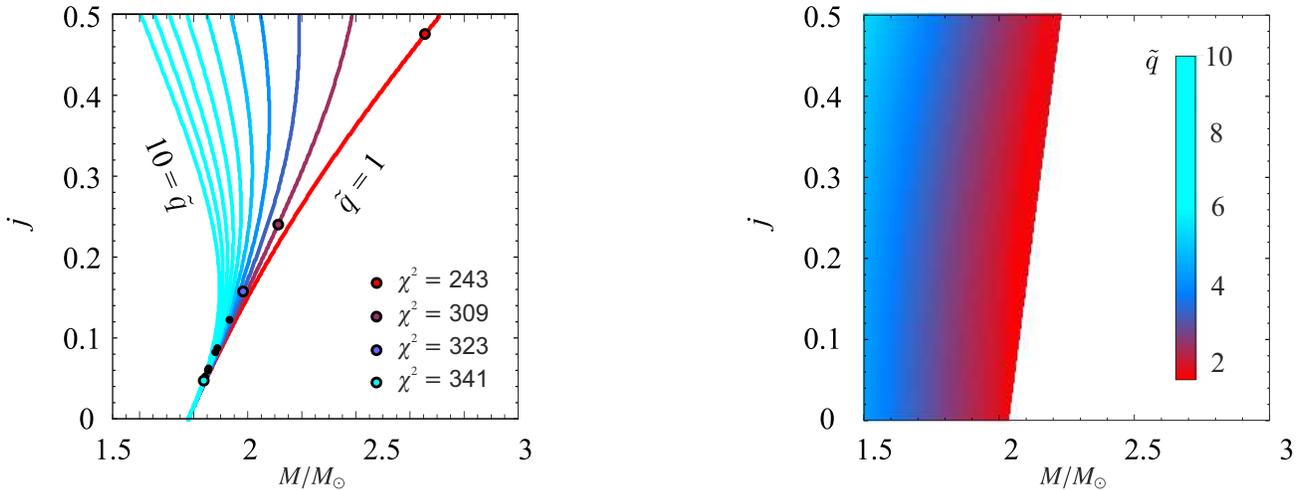}
\end{center}
\caption{Left: Behaviour of the best $\chi^2$ as a function of $M$ and $j$ for several values of $\tilde{q}$. The dots denote global minima for each value of $\tilde{q}$ (see, however, the main text - Section~\ref{section:HTRP}, for a comment on this). Right: The 2D surface in the 3D $M-j-\tilde{q}$ space given by the SLy4  EoS.}
\label{figure:3}
\end{figure*}

Assuming the above formulae we have calculated 3D-$\chi^2$ maps for the RP model. In the left panel of Figure~\ref{figure:3} we show behaviour of the best $\chi^2$ as a function of $M$ and $j$ for several color-coded values of $\tilde{q}$. For each value of $\tilde{q}$ there is a preferred $M-j$ relation. We find that, although such a relation has a global minimum, the gradient of $\chi^2$ along the relation is always much lower than the gradient in the perpendicular direction. In other words, $\chi^2$ maps for a fixed $\tilde{q}$ are of the same type as the one calculated in the Kerr spacetime. It then follows that there is a global $M-j-{\tilde{q}}$ degeneracy in the sense discussed by \cite{tor-etal:2012} -- see their Figure~3.

As emphasized by \cite{urb-etal:2010:b}, \cite{tor-etal:2010}, \cite{Klu-Ros:2013}, \cite{tor-etal:2014}, \cite{Ros-Klu:2014} and \cite{bok-etal:2015}, Newtonian effects following from the influence of the quadrupole moment act on the orbital frequencies in a way opposite to that which is related to relativistic effects following from the increase of the angular momentum. The behaviour of the relations shown in the left panel of Figure~\ref{figure:3} is determined by this interplay. Because of this, we can see that the increased {NS quadrupole moment} can compensate the increase of the estimated mass given by a high angular momentum.

\section{Consideration of NS Models Given by Concrete EoS}
\label{section:EoS}

The relations for the RP model drawn in the left panel of Figure~\ref{figure:3} result from fitting of the 4U 1636-53 datapoints considering the general Hartle-Thorne spacetime. The consideration does not include strong restrictions on spacetime properties following from NS modeling based on present EoS. It can be shown that a concrete NS EoS covers only a 2D surface in the 3D $M-j-\tilde{q}$ space since the quadrupole moment is determined by rotationally induced NS oblateness. Thus, when a given EoS is assumed, only corresponding 2D surface is relevant for fitting of datapoints by a given QPO model. Following \cite{urb-etal:2013}, we illustrate such a surface in the right panel of Figure~\ref{figure:3} for the SLy4~EoS. The color-coding of the plot is the same as the one in the left panel of the same Figure.

\begin{figure*}[t]
\begin{center}
\includegraphics[width=1\linewidth]{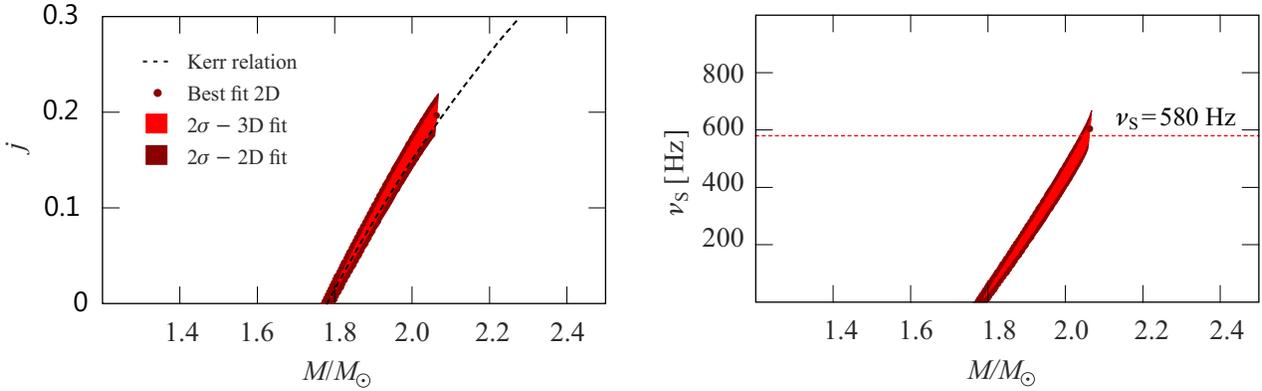}
\end{center}
\caption{Left: The final $M-j$ map implied by the RP model and the Sly4 EoS.  The light colour area denotes an intersection between the 2D surface given by the Sly4 EoS and 3D volume corresponding to $2\sigma$ confidence level  given by the RP model best fit found in the intervals of $M\in [1M_{\sun},\,4M_{\sun}]$, $j \in [0.0,0.5]$, and $\tilde{q} \in [1,10]$. The $\chi^2$ minimum of {$303/20$d.o.f.} at the 2D surface is denoted by the dark marker. The dark colour area denotes the $2\sigma$ confidence level calculated when  this local minimum is assumed as a global one providing that the QPO frequency error bars are underestimated by a corresponding factor $\xi_{\mathrm{2D}} = 3.9$. We can see that,  in this particular case, there is almost no difference between the two areas. The dashed curve indicates the $M-j$ relation obtained from the simplified consideration of Kerr spacetimes (see Section~\ref{section:fits:Kerr}).  Right: The same map, but drawn for the NS spin frequency $\nu_{\mathrm{S}}$. The horizontal dashed red line denotes the  spin frequency measured from the X-ray bursts (i.e., $\nu_{\mathrm{S}}=580$Hz). The minimum of $\chi^2$ for the spin  $\nu_{\mathrm{S}}=580$Hz corresponds to $\chi^2 = 305/21$d.o.f. }
\label{figure:4}
\end{figure*}

The final $M-j$ map for the RP model and Sly4 EoS, i.e.,  the values of $M$ and $j$ implied by the common consideration of both panels of Figure~\ref{figure:3}, is shown in the left panel of Figure~\ref{figure:4}. The right panel of this Figure then shows an equivalent map drawn for the NS mass and spin frequency $\nu_{\mathrm{S}}$. 

\begin{figure*}[t]
	\begin{center}
		\includegraphics[width=1\linewidth]{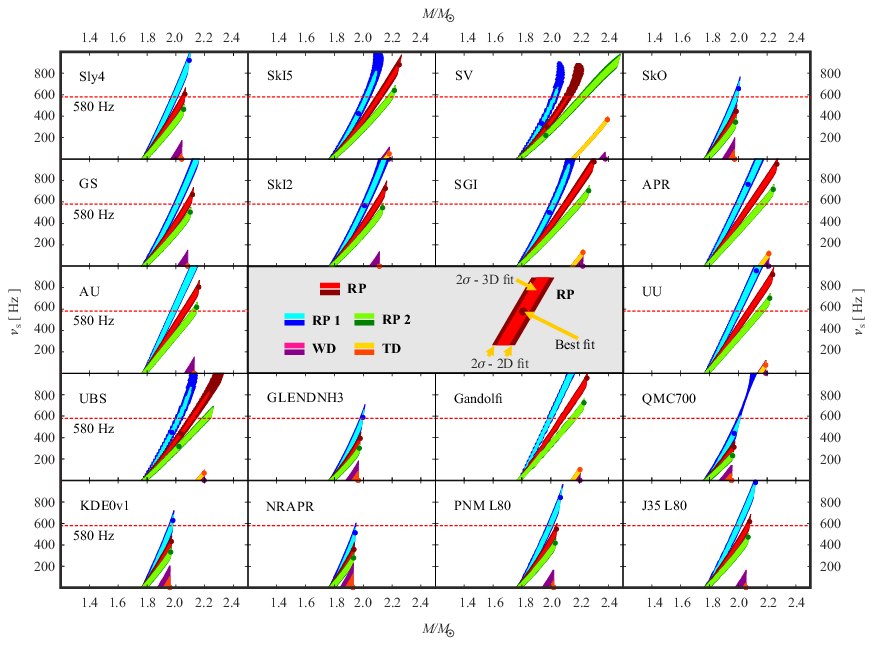}
	\end{center}
	\caption{The mass-spin maps for the considered QPO models and 18 different EoS. The light colour area denotes an intersection between the 2D surface given by the EoS and 3D volume corresponding to $2\sigma$ confidence level given by the  model best fit found in the intervals of $M\in [1M_{\sun},\,4M_{\sun}]$, $j \in [0.0,0.5]$, and $\tilde{q} \in [1,10]$. The $\chi^2$ minimum at the 2D surface is denoted by the dark marker. The dark colour area denotes the $2\sigma$ confidence level calculated when this local minimum is assumed as a global one providing that the QPO frequency error bars are underestimated by a corresponding factor $\xi_{\mathrm{2D}}$.  The horizontal dashed red lines denote spin frequency measured from the X-ray bursts (i.e., $\nu_{\mathrm{S}}=580$Hz).}
	\label{figure:5}
\end{figure*}

\subsection{NS Mass Inferred Assuming X-ray Burst Measurements}
\label{section:spin}

Inspecting the left panel of Figure~\ref{figure:4}, we can see that the concrete EoS, SLy4, considered for the RP model implies a clear $M-j$ relation. This relation exhibits only a shallow $\chi^2$ minimum. The right panel of the same Figure shows the equivalent relation between the NS mass and the spin frequency as well as its shallow minimum. Taking into account the  spin frequency inferred from the X-ray bursts, 580Hz, we can find from Figure~\ref{figure:4} that the NS mass and angular momentum have to take values,
\begin{equation}
M=(2.06\pm0.01)M_\odot\,,\quad j\doteq0.2\,.
\end{equation}
These values are in a good agreement with those inferred from the simplified consideration using Kerr spacetimes (see Figure~\ref{figure:2}). Considering the shallow $\chi^2$ minima denoted in Figure~\ref{figure:4}, it may be interesting that its frequency value almost coincides with the measured spin frequency of 580Hz.

\section{Discussion and Conclusions}\label{section:conclusions}



As well as the Sly4 EoS, we have investigated a wide set of 17 other EoS that are based on different theoretical models. All these EoS are listed in Table~\ref{table:1} where we show the maximum NS mass allowed by each EoS as well as the corresponding NS radius and the central number density. All these EoS are compatible with the highest observed NS masses (see, e.g., \cite{klaehn}, \cite{ste-lat-bro}, \cite{ste-gan-fat-new}, \cite{klaehn2}, \cite{dutra1} and \cite{dutra2}  for various tests of EoS and their applications, and \cite{demorest} and \cite{antoniadis:2013} for the highest observed NS masses). 

In Figure~\ref{figure:5} we show several relations between the mass and spin frequency obtained for the RP model and our large set of EoS. These relations are similar to those implied by the Sly4 EoS discussed above. However, we can see that, in several cases, a given EoS does not provide any match for the NS spin of $580$Hz. This can rule out the combination of the considered RP model and given specific EoS. The selection effect comes from the correlation between the estimated mass and angular momentum and the limits on maximal mass allowed by the individual EoS.

\subsection{Selecting Combinations of QPO Models and EoS}

We have found an analogic selection effect also for the other four examined QPO models. The corresponding $M-j$ maps are shown in Figure~\ref{figure:5}. The results for all considered  models are summarized in Table~\ref{table:2}. In the table, we can find which of the models and EoS are compatible, and which of them are not. Overall, there are  39 matches from the 90 investigated cases for the  NS spin frequency of $580$Hz. We can therefore conclude that, for the NS spin frequency in 4U~1636-53 to be close to 580Hz, we can exclude 51 from the 90 considered combinations of EoS and QPO models. This result follows from the requirement of relatively large masses implied by the individual QPO models and increase of these masses with the NS spin.



\begin{table*}\scriptsize
\caption{Results for the considered EoS and QPO models.  \label{table:2}}
\begin{center}
{\begin{tabular}{cccccccccc}\tableline \tableline \\
&  \multicolumn{3}{ c }{RP model }  & \multicolumn{3}{ c }{RP1 model } & \multicolumn{3}{ c }{RP2 model }  \\
 & \multicolumn{3}{ c }{$\xi_{H+T}$ = 3.6, $\xi_{Kerr}$ = 3.1}  & \multicolumn{3}{ c }{ $\xi_{H+T}$ = 3.6, $\xi_{Kerr}$ = 3.3} & \multicolumn{3}{ c }{ $\xi_{H+T}$ = 3.9, $\xi_{Kerr}$ = 3.5} \\ 
EoS    & $M$ &$j$ & $\chi^2_{min}$ & $M$ & $j$ & $\chi^2_{min}$ & $M$ & $j$ & $\chi^2_{min}$ \\ \\ \tableline 
  SLy4  & 2.06 $\pm$ 0.01   &0.19  & 305 & 1.99 $\pm$ 0.02  & 0.21  & 302 & X & X  & X \\  
	SkI5   & 2.11 $\pm$ 0.01   &0.25  & 323 & 2.02 $\pm$ 0.01  & 0.27  & 321 & 2.19 $\pm$ 0.02  & 0.23   & 327 \\ 
	SV   & 2.12 $\pm$ 0.01   &0.28  & 345 & 2.03 $\pm$ 0.01  & 0.30  & 334 & 2.22 $\pm$ 0.02  & 0.27   & 355 \\  
	SkO  & X  &X & X & 1.98 $\pm$ 0.01  & 0.20  & 302 & X & X  & X \\ 
  Gs  & 2.08 $\pm$ 0.01   &0.22  & 309 & 2.01 $\pm$ 0.02  & 0.24  & 307 & X & X  & X \\  
  SkI2  & 2.10 $\pm$ 0.01   &0.23  & 313 & 2.01 $\pm$ 0.02  & 0.25  & 311 & 2.14 $\pm$ 0.01  & 0.21   & 395 \\   
  SGI   & 2.11 $\pm$ 0.01   &0.25  & 319 & 2.02 $\pm$ 0.02  & 0.26  & 314 & 2.19 $\pm$ 0.02  & 0.23   & 328 \\   
  APR  & 2.09 $\pm$ 0.01   &0.22  & 309 & 2.00 $\pm$ 0.02  & 0.23  & 304 & 2.17 $\pm$ 0.02  & 0.21   & 320 \\  
  AU  & 2.06 $\pm$ 0.01   &0.20  & 305 & 1.98 $\pm$ 0.02  & 0.20  & 301 & 2.13 $\pm$ 0.02  & 0.19   & 315 \\  
  UU  & 2.08 $\pm$ 0.01   &0.21  & 306 & 1.99 $\pm$ 0.02  & 0.22  & 302 & 2.16 $\pm$ 0.02  & 0.20   & 317 \\ 
	UBS   & 2.11 $\pm$ 0.01   &0.26  & 325 & 2.02 $\pm$ 0.01  & 0.27  & 317 & 2.21 $\pm$ 0.02  & 0.25   & 338 \\ 
	GLENDNH3  & X  &X & X & 2.00 $\pm$ 0.01  & 0.22  & 303 & X & X  & X \\   
  Gandolfi  & 2.08 $\pm$ 0.01   &0.21  & 307 & 1.99 $\pm$ 0.01  & 0.22  & 303 & 2.15 $\pm$ 0.02  & 0.20   & 318 \\  
  QMC700  & X  &X & X & 2.01 $\pm$ 0.01  & 0.27  & 332 & X & X  & X \\  
  KDE0v1  & X  &X & X & 1.97 $\pm$ 0.01  & 0.19  & 301 & X & X  & X \\  
  NRAPR  & X  &X & X & 1.95 $\pm$ 0.01* & 0.18* & 7196 & X & X  & X \\   
  PNM L80  & 2.04 $\pm$ 0.01   &0.19  & 340 & 2.00 $\pm$ 0.01  & 0.22  & 303 & X & X  & X \\   
  J35 L80   & 2.07 $\pm$ 0.01   &0.21  & 306 & 2.00 $\pm$ 0.01  & 0.23  & 304 & X & X  & X \\  \tableline
	\end{tabular}}
\tablecomments{ Values of $\xi = \sqrt{\chi_{min}^{2}/dof}$ corresponding to global minima of $\chi_{min}^{2}$ in Hartle-Thorne spacetime are compared to values obtained for Kerr spacetimes in \cite{tor-etal:2012}. Asterisks denote that the indicated values of $M$ and $j$ are connected to the dark colour area in Figures \ref{figure:4} and \ref{figure:5}. In such case there is no intersection between the spin frequency curve in the 2D EoS plane and $2\sigma$ level 3D volume around the global minima of the QPO model fit in the Hartle-Thorne spacetime. The local minimum, $\chi_{{\mathrm{2Dmin}}}^{2}$, at the 2D EoS plane is than assumed as a global one providing that the QPO frequency error bars are underestimated by a corresponding factor, $\xi_{\mathrm{2D}} = \sqrt{\chi_{{\mathrm{2Dmin}}}^{2}/dof}$. The $X$-symbol indicates that the spin frequency $580$Hz is not reached even in this case. {For the TD and WD models, the spin frequency is not reached.}}

\end{center}

\end{table*}

\begin{figure*}[t]
\begin{center}
\includegraphics[width=.95\linewidth]{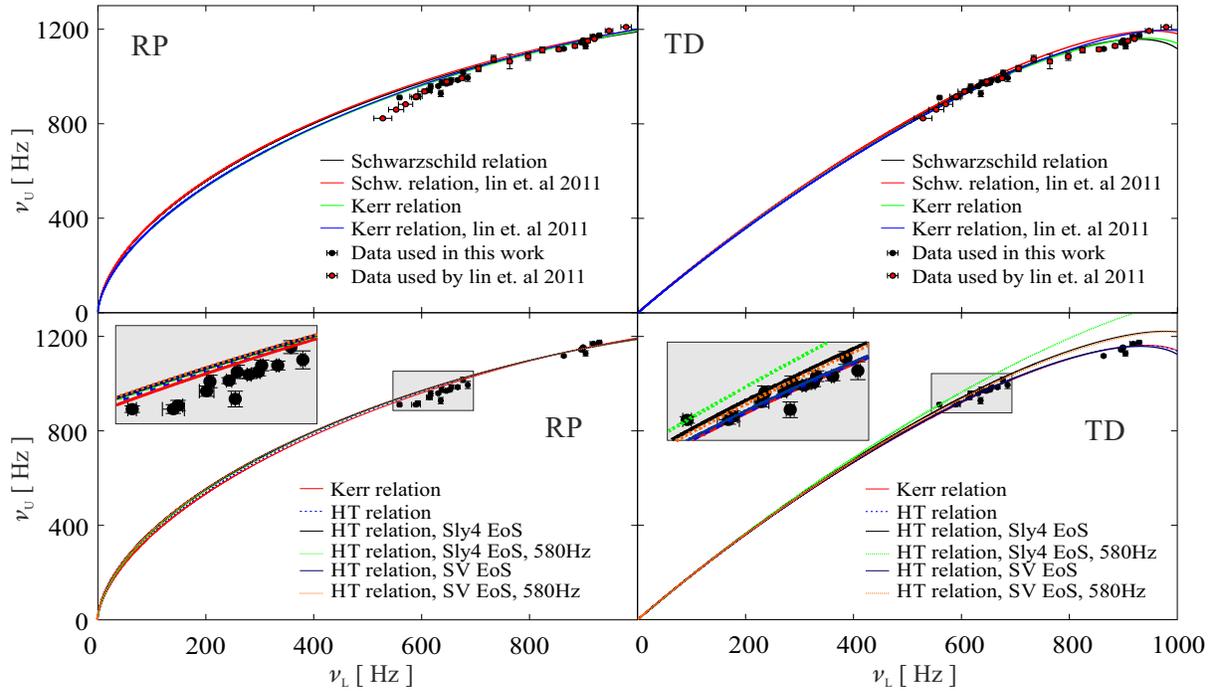}
\end{center}
\caption{The twin peak QPO data and examples of their individual fits. Top: The best fit of data (black dots) used in this work and the best fit of data (black-red dots) used by \cite{Lin-etal:2011}. Bottom: Best fits assuming Kerr spacetime denoted by red lines ($\chi^2_{min,RP} = 189/20$d.o.f., $\chi^2_{min,TD} = 122/20$d.o.f). Best fits in Hartle-Thorne spacetimes denoted by blue lines ($\chi^2_{min,RP} = 243/19$d.o.f, $\chi^2_{min,TD} = 123/19$d.o.f.). Best fits in Hartle-Thorne spacetimes restricted to the parametric 2D surface given by SLy 4 EoS denoted by black lines ($\chi^2_{min,RP} = 303/20$d.o.f, $\chi^2_{min,TD} = 2514/20$d.o.f). Best fits in Hartle-Thorne spacetimes restricted to the parametric 2D surface given by SV EoS denoted by dark blue lines ($\chi^2_{min,RP} = 327/20$d.o.f, $\chi^2_{min,TD} = 129/20$d.o.f). Best fits in Hartle-Thorne spacetimes restricted by SLy 4 Eos and NS spin $580$Hz denoted by the green lines ($\chi^2_{min,RP} = 305/21$d.o.f, $\chi^2_{min,TD} = 15725/21$d.o.f). Best fits in Hartle-Thorne spacetimes restricted by  SV EoS and NS spin $580$Hz denoted by the orange lines ($\chi^2_{min,RP} = 344/21$d.o.f, $\chi^2_{min,TD} = 2233/21$d.o.f).
}
\label{figure:6}
\end{figure*}

\begin{figure*}[h!]
\begin{center}
\includegraphics[width=1.\linewidth]{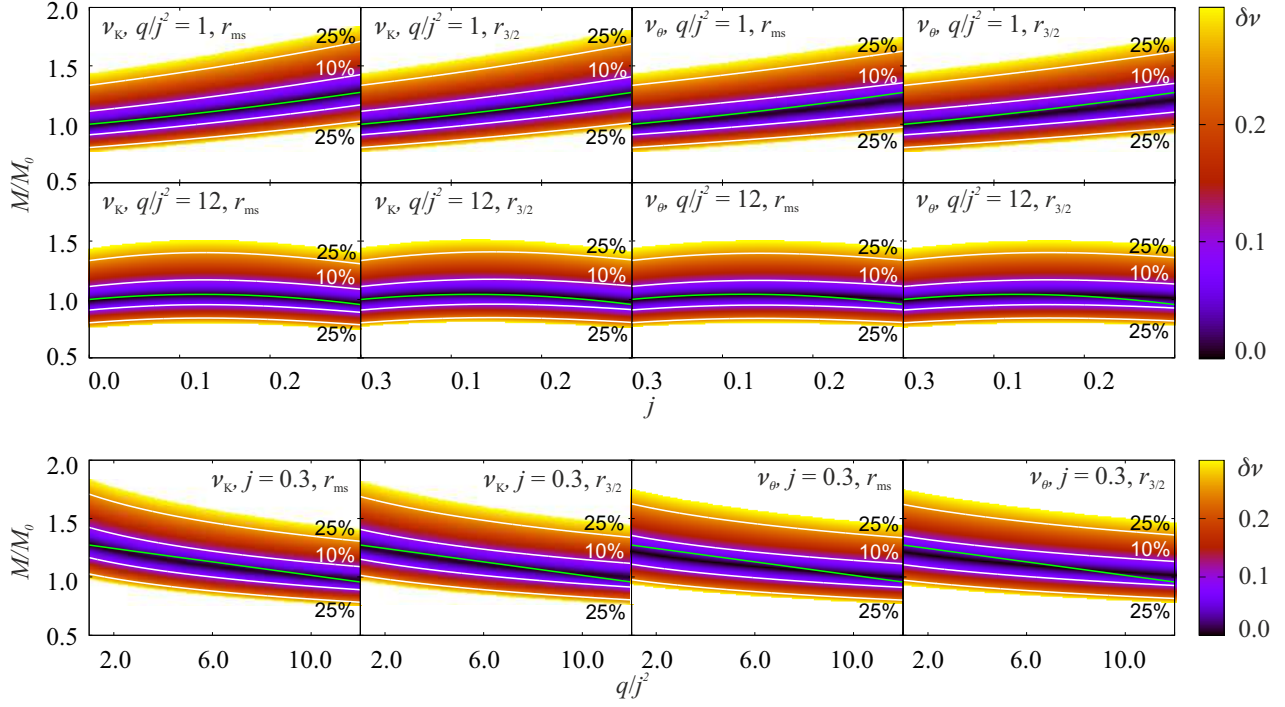}
\end{center}
\caption{Color-coded maps indicating values of dimensionless difference between characteristic frequencies of orbital motion $\delta \nu\equiv(\nu(M_0)-\nu(M,\,j,\,q))/\nu(M_0)$ calculated assuming the  Schwarzschild spacetimes $(M=M_{0})$ and Hartle--Thorne spacetimes. Individual panels assume chosen fixed values of  parameters $j$ and~$\tilde{q}$. Combinations of parameters indicated by green curves are given by relation (\ref{equation:HTdeg}), $M = M_0 \left(1+0.7j+1.02j^2-0.32 q\right)$.  Frequencies are calculated at characteristic radii $r_{3:2}$ and $r_{\mathrm{ms}}$ where the Keplerian and periastron precession frequencies are in a 3:2 and 1:1 ratio.}
\label{figure:7}
\end{figure*}

\begin{figure*}[ht]
\begin{center}
\includegraphics[width=.55\linewidth]{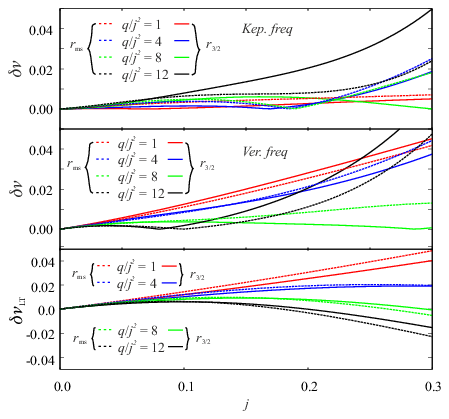}
\end{center}
\caption{Dimensionless quantity $\delta \nu$ plotted for different values of $q/j^2$ and  relation (\ref{equation:HTdeg}). We do not include here panels for radial and periastron precession frequencies because the values are the same as for the Keplerian frequency.}
\label{figure:8}
\end{figure*}

\subsection{Implications for QPO Models}


Assuming the Hartle-Thorne geometry restricted to the range of angular momentum and {scaled quadrupole moment} $\{j,\tilde{q}\}\in{\{[0,\,0.4],[1,\,10]\}}$, the four investigated QPO models imply a relatively large range of NS mass, $M\in[1.6,\,3.4]\,M_{\sun}$ ($M\in[1.8,\,2.5]\,M_{\sun}$ when $j=0$). In Figure~\ref{figure:6} we illustrate a corresponding comparison between the data and some individual fits. Inspecting Figure~\ref{figure:6} we can see that the quality of fits is rather poor (represented by $\chi^2/\mathrm{d.o.f.}\sim10$, see Table~\ref{table:2}). The comparison between data and curves drawn for the RP model indicates possible presence of systematic errors within the model. This is also valid for the RP1, RP2 and WD model. The trend is somewhat better only in the case of the TD model. This has been noticed also by \cite{Lin-etal:2011}.  However, when we take into account requirements given by present EoS and the NS spin of $580$Hz, the TD model is ruled out (see the green curve in the bottom right panel of Figure~\ref{figure:6}). The range of NS mass corresponding to considered models is then reduced to $M\in[2.0,\,2.2]\,M_{\sun}$.

Remarkably, the consideration of Hartle-Thorne spacetime does not improve the quality of fits. For instance, the deviation of the RP model curve from the data discussed  by \cite{Lin-etal:2011} is present when we assume Hartle-Thorne as well as Kerr spacetime. There is  $\Delta{\chi}\equiv\sum\mathrm{sign}(\chi_{\mathrm{i}})\chi^2_{\mathrm{i}}\sim -150$ for the bottom part of the curve (i$\,\in\{1\ldots 14\}$) while it is $\Delta{\chi}\sim +20$ for the top part of the curve (i$\,\in\{15\ldots 22\}$). Possible need of non-geodesic corrections discussed by \cite{tor-etal:2012} and \cite{Lin-etal:2011} therefore does not depend on the chosen spacetime description \citep[see also][in this context]{torok-etal:2015:MNRAS:}. This conclusion is in a good agreement with the suggestion of \cite{tor-etal:2012} implying that parameters of RP model fits within Hartle-Thorne spacetime should exhibit a degeneracy approximated as
\begin{equation}
	\label{equation:HTdeg}
	M = M_0 \left(1+0.7j+1.02j^2-0.32 q\right),
\end{equation}
where $M_{0}=1.78M_{\sun}$ for 4U~1636-53. This degeneracy is illustrated in Figures \ref{figure:7} and \ref{figure:8} where we also quantify its validity for the other models discussed here.

\begin{figure*}[t]
\begin{center}
\includegraphics[width=.95\linewidth]{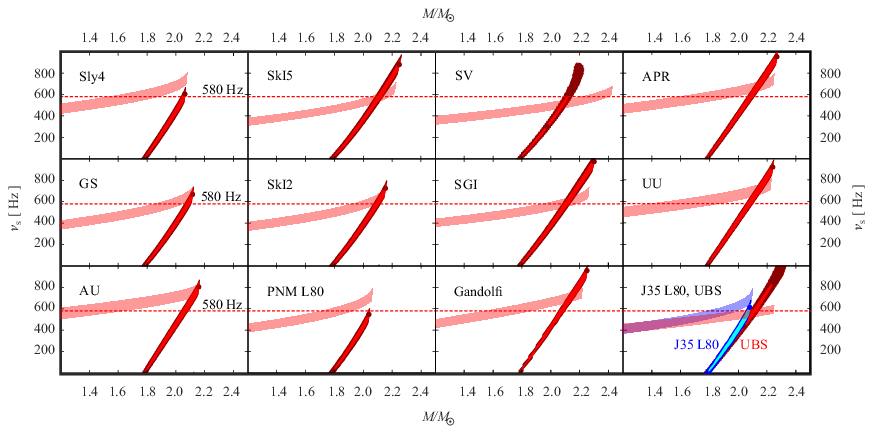}
\end{center}
\caption{Consideration of RP model assuming both low and high frequency QPOs and 13 EoS. The RP model mass-spin maps from Figure~\ref{figure:5} are confronted with requirements following from the identification of low frequency QPOS with the Lense-Thirring precession frequency. The last panel includes the consideration of two different EoS.}
\label{figure:9}
\end{figure*}

\subsection{Consideration of Low Frequency QPOs}

Strong restrictions to the model and implied NS mass may be obtained when low frequency QPOs are considered. This can be clearly illustrated for the RP model which associates the observed low-frequency QPOs to the Lense--Thirring precession that occurs at the same radii as the periastron precession. Within the framework of the  model,  the Lense--Thirring frequency $\nuLT$ represents  a  sensitive  spin  indicator  \citep[][]{ste-vie:1998:a, ste-vie:1998:b, Mor-ste:1999:,ste-etal:1999}. In our previous paper \citep{tor-etal:2012} we carried out a simplified estimate of the underlying NS angular momentum and mass assuming Kerr spacetimes, arriving at the values of $j\sim 0.25\div0.35$ and $M\sim (2.2\div2.4)M_{\sun}$. These values appeared too high when confronted with the implications of the set of 5 EoS assumed within the paper. As discussed here in Section~\ref{section:fits:Kerr}, the extended set of EoS can be more compatible with the expectations based on the consideration of Lense--Thirring precession. It is straightforward to extend our previous estimate to Hartle-Thorne spacetime and all 18 EoS. Results of such extension are included in Figure~\ref{figure:9}. We show there {13} EoS compatible with the observed twin peak QPOs and RP model, and demonstrate that 8 of these EoS do not meet requirements based on the consideration of Lense-Thirring preccession. Only {5} EoS are thus compatible with the model.

\bigskip
\acknowledgments
\noindent{ACKNOWLEDGMENTS}
\medskip

{We would like to acknowledge the Czech grant GA\v{C}R 209/12/P740, and internal grants of the Silesian University in Opava, SGS/11,23/2013, SGS/14,15/2016 and IGS/12/2015. {ZS acknowledges the Albert Einstein Center for Gravitation and Astrophysics supported by the Czech Science Foundation grant No. 14-37086G}. We are grateful to Marek Abramowicz, Wlodek Kluzniak (CAMK), John Miller (University of Oxford), Will Newton (Texas A\&M University-Commerce), Luigi Stella (INAF), and Ji\v{r}ina Stone (Oak Ridge National Laboratory) for many useful discussions. {We also thank to the anonymous referee for his/her comments and suggestions that greatly helped to improve the paper.} Furthermore we would like to acknowledge the hospitality of the University of Oxford and the Astronomical Observatory in Rome. {Last but not least, we express our sincere thanks to concierges of Ml\'{y}nsk\'{a} hotel in Uhersk\'{e} Hradi\v{s}t\v{e} for their kind help and participation in organizing frequent workshops of the Silesian university and the Astronomical institute.}}
\bigskip\bigskip



\begin{thebibliography}{}
\bibitem[Abramowicz et al.(2003a)]{abr-etal:2003a} {Abramowicz}, M.~A., {Almergren}, G.~J.~E., {Klu{\'z}niak}, W., \& {Thampan}, A.~V. 2003a, ArXiv General Relativity and Quantum Cosmology e-prints, gr-qc/0312070
\bibitem[Abramowicz et~al.(2003b)]{abr-etal:2003b}  {Abramowicz}, M.~A., {Bulik}, T., {Bursa}, M., \& {Klu{\'z}niak}, W. 2003b, A\&A, 404, L21
\bibitem[Abramowicz et~al.(2003c)]{abr-etal:2003c} {Abramowicz}, M.~A., {Karas}, V., {Klu{\'z}niak}, W., {Lee}, W.~H., \& {Rebusco}, P. 2003c, PASJ,  55, 467
\bibitem[Abramowicz \& {Klu{\'z}niak}(2001)]{abr-klu:2001} {Abramowicz}, M.~A., \& {Klu{\'z}niak}, W. 2001, A\&A, 374, L19
\bibitem[Agrawal et~al.(2005)]{kde0v1} {Agrawal}, B.~K., {Shlomo}, S., \& {Au}, V.~K. 2005, PhRvC, 72, 014310
\bibitem[Akmal et al.(1998)]{akm-etal:1998} {Akmal}, A., {Pandharipande}, V.~R., \& {Ravenhall}, D.~G. 1998, PhRvC, 58, 1804
\bibitem[Aliev \& Galtsov(1981)]{ali-gal:1981} {Aliev}, A.~N., \& {Galtsov}, D.~V. 1981, GReGr, 13, 899
\bibitem[Alpar \& Shaham(1985)]{alp-sha:1985} {Alpar}, M.~A., \& {Shaham}, J. 1985, Natur, 316, 239
\bibitem[Ansorg et al(2003)]{ans-etal:2003} {Ansorg}, M., {Kleinw{\"a}chter}, A. \& {Meinel}, R. 2003, A\&A, 405, 711
\bibitem[Antoniadis et~al.(2013)]{antoniadis:2013} {Antoniadis}, J., {Freire}, P.~C.~C., {Wex}, N., {et~al.} 2013,  Sci, 340, 448
\bibitem[Barret \& Boutelier(2008)]{bar-bou:2008} {Barret}, D., \& {Boutelier}, M. 2008, NewAR, 51, 835
\bibitem[Barret et al.(2005)]{bar-etal:2005} {Barret}, D., {Olive}, J.-F., \& {Miller}, M.~C. 2005, MNRAS, 361, 855
\bibitem[Barret et al.(2006)]{bar-etal:2006} {Barret}, D., {Olive}, J.-F., \& {Miller}, M.~C. 2006, MNRAS, 370, 1140
\bibitem[Belloni et al.(2007)]{bel-etal:2007} {Belloni}, T., {Homan}, J., {Motta}, S., {Ratti}, E., \& {M{\'e}ndez}, M. 2007, MNRAS, 379, 247
\bibitem[Belloni et al.(2005)]{bel-etal:2005} {Belloni}, T.,  {M{\'e}ndez}, M., \& {Homan}, J. 2005, A\&A, 437, 209
\bibitem[Berti et al.(2005)]{ber-etal:2005} {Berti}, E., {White}, F., {Maniopoulou}, A. \& {Bruni}, M. 2005, MNRAS, 358, 923
\bibitem[Bonazzola et al.(1993)]{bon-etal:1993} {Bonazzola}, S., {Gourgoulhon}, E., {Salgado}, M. \& {Marck}, J.~A. 1993, A\&A, 278, 421 
\bibitem[Bonazzola et al.(1998)]{bon-etal:1998} {Bonazzola}, S., {Gourgoulhon}, E. \& {Marck}, J.-A. 1998, PrD, 58, 104020
\bibitem[{{Boshkayev} {et~al.}(2015)}]{bok-etal:2015} {Boshkayev}, K., {Quevedo}, H., {Abutalip}, M., {Kalymova}, Z., \& {Suleymanova}, S. 2015, eprint arXiv:1510.02016
\bibitem[Boutelier et al.(2010)]{bou-etal:2010} {Boutelier}, M., {Barret}, D., {Lin}, Y., \& {T{\"o}r{\"o}k}, G. 2010, MNRAS, 401, 1290
\bibitem[Bursa(2005)]{bur:2005} {Bursa}, M. 2005, in RAGtime 6/7: Workshops on black holes and neutron stars, ed. S. {Hled{\'{\i}}k} \& Z. {Stuchl{\'{\i}}k}, 39–45
\bibitem[{{\v C}ade{\v z}} et al.(2008)]{cad-etal:2008}  {{\v C}ade{\v z}}, A., {Calvani}, M., \& {Kosti{\'c}}, U. 2008, A\&A, 487, 527
\bibitem[Chandrasekhar \& Miller(1974)]{cha-mil:1974} {Chandrasekhar}, S., \& {Miller}, J.~C. 1974,  MNRAS, 167, 63
\bibitem[{Demorest} et~al.(2010)]{demorest} {Demorest}, P.~B., {Pennucci}, T., {Ransom}, S.~M., {Roberts}, M.~S.~E., \& {Hessels}, J.~W.~T. 2010, Natur, 467, 1081
\bibitem[{Dutra} {et~al.}(2012)]{dutra1} {Dutra}, M., {Louren{{ c}}o}, O., {S{\'a} Martins}, J.~S., {et~al.} 2012, Physical Review C: Nuclear Physics, 85, 035201
\bibitem[{Dutra} {et~al.}(2014)]{dutra2} {Dutra}, M., {Louren{ c}o}, O., {Avancini}, S.~S., {et~al.} 2014, Physical Review C: Nuclear Physics, 90, 055203
\bibitem[{Gandolfi} {et~al.}(2010)]{gandolfi} {Gandolfi}, S., {Illarionov}, A.~Y., {Fantoni}, S., {et~al.} 2010, MNRAS, 404, L35
\bibitem[Galloway et al.(2008)]{gal-etal:2008:APJS:} Galloway, D.~K., Muno, M.~P., Hartman, J.~M., Psaltis, D., \& Chakrabarty, D. 2008, \apjs, 179, 360
\bibitem[{German{\`a}} et al.(2009)]{ger-etal:2009} {German{\`a}}, C., {Kosti{\'c}}, U., {{\v C}ade{\v z}}, A., \& {Calvani}, M. 2009,  in American Institute of Physics
Conference Series, Vol. 1126, American Institute of Physics Conference Series, ed.
J. Rodriguez \& P. Ferrando, 367-369
\bibitem[Gilfanov et al.(2000)]{gilf-etal:2000} {Gilfanov}, M., {Churazov}, E., \& {Revnivtsev}, M. 2000, MNRAS, 316, 923
\bibitem[{{Glendenning}(1985)}]{glendnh3} {Glendenning}, N.~K. 1985,  ApJ, 293, 470
\bibitem[Hartle(1967)]{har:1967} {Hartle}, J.~B. 1967, ApJ, 150, 1005    
\bibitem[{Hartle} \& {Thorne}(1968)]{har-tho:1968} {Hartle}, J.~B., \& {Thorne}, K.~S. 1968, ApJ, 153, 807
\bibitem[Hor{\'a}k et al.(2009)]{hor-etal:2009} {Hor{\'a}k}, J.,  {Abramowicz}, M.~A., {Klu{\'z}niak}, W., {Rebusco}, P., \& {T{\"o}r{\"o}k}, G. 2009, A\&A, 499, 535
\bibitem[{Jonker et al.}(2005)]{jon-etal:2005:} Jonker, P. G., M\'endez, M. \& van der Klis, M., 2005, MNRAS, 360, 3, 921
\bibitem[Kato(2001)]{kat:2001} {Kato}, S. 2001, PASJ, 53, 1
\bibitem[Kato(2007)]{kat:2007} {Kato}, S. 2007, PASJ, 59, 451
\bibitem[Kato(2008)]{kat:2008} {Kato}, S. 2008, PASJ, 60, 111
\bibitem[{Kl{\"a}hn} {et~al.}(2007)]{klaehn2} {Kl{\"a}hn}, T., {Blaschke}, D., {Sandin}, F., {et~al.} 2007,  PhLB, 654, 170
\bibitem[{Kl{\"a}hn} {et~al.}(2006)]{klaehn}{Kl{\"a}hn}, T., {Blaschke}, D., {Typel}, S., {et~al.} 2006, Physical Review C:Nuclear Physics, 74, 035802
\bibitem[Klu{\'z}niak \& {Abramowicz}(2001)]{klu-abr:2001} {Klu{\'z}niak}, W., \& {Abramowicz}, M.~A. 2001, ArXiv Astrophysics e-prints, astro-ph/0105057
\bibitem[Klu{\'z}niak \& {Abramowicz}(2002)]{klu-abr:2002} {Klu{\'z}niak}, W., \& {Abramowicz}, M.~A. 2001,  ArXiv Astrophysics e-prints, astro-ph/0203314
\bibitem[Klu{\'z}niak et al.(2004)]{klu-etal:2004} {Klu{\'z}niak}, W., {Abramowicz}, M.~A., {Kato}, S., {Lee}, W.~H., \& {Stergioulas}, N. 2004, ApJ, 603, L89
\bibitem[{Klu{\'z}niak} \& {Rosi{\'n}ska}(2013)]{Klu-Ros:2013} {Klu{\'z}niak}, W., \& {Rosi{\'n}ska}, D. 2013, MNRAS, 434, 2825
\bibitem[Kosti{\'c} et al.(2009)]{kos-etal:2009} {Kosti{\'c}}, U.,  {{\v C}ade{\v z}}, A.,  {Calvani}, M., \& {Gomboc}, A. 2009, A\&A, 496, 307
\bibitem[Lamb  et al.(1985) ]{lam-etal:1985} {Lamb}, F.~K., {Shibazaki}, N., {Alpar}, M.~A., \& {Shaham}, J. 1985, Natur, 317, 681
\bibitem[{{Lin} {et~al.}(2011)}]{Lin-etal:2011}{Lin}, Y.-F., {Boutelier}, M., {Barret}, D., \& {Zhang}, S.-N. 2011,  ApJ, 726, 74
\bibitem[Manko et al.(2000)]{Manko-etal:2000:} {Manko}, V.~S., {Mielke}, E.~W. \& {Sanabria-G{\'o}mez}, J.~D. 2000, PrD, 61, 081501
\bibitem[{M{\'e}ndez}(2006)]{men:2006} {M{\'e}ndez}, M. 2006, MNRAS, 371, 1925
\bibitem[{Miller}(1977)]{mil:1977} {Miller}, J.~C. 1977, MNRAS, 179, 483
\bibitem[Miller et al.(1998)]{mil-etal:1998a} {Miller}, M.~C., {Lamb}, F.~K., \& {Psaltis}, D. 1998, ApJ, 508, 791
\bibitem[Morsink \& Stella(1999)] {Mor-ste:1999:} Morsink, S. M., \& Stella, L. 1999, ApJ, 513, 827
\bibitem[Mukhopadhyay(2009)]{muk:2009} {Mukhopadhyay}, B. 2009, ApJ, 694, 387 
\bibitem[{{Newton} {et~al.}(2013)}]{will} {Newton}, W.~G., {Gearheart}, M., \& {Li}, B.-A. 2013,  ApJS, 204, 9
\bibitem[Nozawa {et~al.}(1998)]{noz-etal:1998} {Nozawa}, T., {Stergioulas}, N., {Gourgoulhon}, E. \& {Eriguchi}, Y. 1998, A \& A Sup., 132,431
\bibitem[Pappas(2015)]{pap:2015} {Pappas}, G. 2015, MNRAS, 454, 4066
\bibitem[P{\'e}tri(2005)]{pet:2005a} {P{\'e}tri}, J. 2005, A\&A, 439, L27
\bibitem[Psaltis et al.(1999)]{psa-etal:1999b} {Psaltis}, D.,  {Wijnands}, R., \& {Homan}, J., et al. 1999, ApJ, 520, 763
\bibitem[Rezzolla et al.(2003)]{rez-etal:2003} {Rezzolla}, L., {Yoshida}, S., \& {Zanotti}, O. 2003, MNRAS, 344, 978
\bibitem[{{Rikovska Stone} {et~al.}(2007)}]{qmc700} {Rikovska Stone}, J., {Guichon}, P.~A.~M., {Matevosyan}, H.~H., \& {Thomas}, A.~W. 2007, NuPhA, 792, 341
\bibitem[Rikovska Stone et al.(2003)]{rik-etal:2003} {Rikovska Stone}, J., {Miller}, J.~C., {Koncewicz}, R., {Stevenson}, P.~D., \& {Strayer}, M.~R. 2003, PhRvC, 68, 034324
\bibitem[Rosi{\'n}ska et al.(2014)]{Ros-Klu:2014} {Rosi{\'n}ska}, D., {Klu{\'z}niak}, W., {Stergioulas}, N., \& {Wi{\'s}niewicz}, M. 2014, PhRvD, 89, 104001  
\bibitem[Stute \& Camenzind(2002)]{stu-gam:2002:} {Stute}, M. \& {Camenzind}, M. 2002, MNRAS, 336, 831-840
\bibitem[{{Steiner} {et~al.}(2015)}]{ste-gan-fat-new} {Steiner}, A.~W., {Gandolfi}, S., {Fattoyev}, F.~J., \& {Newton}, W.~G. 2015, Physical Review C: Nuclear Physics, 91, 015804
\bibitem[{{Steiner} {et~al.}(2010)}]{ste-lat-bro} {Steiner}, A.~W., {Lattimer}, J.~M., \& {Brown}, E.~F. 2010,  ApJ, 722, 33
\bibitem[{{Steiner} {et~al.}(2005)}]{nrapr} {Steiner}, A.~W., {Prakash}, M., {Lattimer}, J.~M., \& {Ellis}, P.~J. 2005,  PhR, 411, 325
\bibitem[{Stella}, L. \& {Vietri}(1999)]{Ste-Vi:1999} {Stella}, L., \& {Vietri}, M. 1999, PhRvL, 82, 17
\bibitem[Stella \& Vietri(1998a)]{ste-vie:1998:a} Stella, L., \& Vietri, M. 1998a, in Abstracts of the 19th Texas Symposium on Relativistic Astrophysics and Cosmology, ed. J. Paul, T. Montmerle, \& E. Aubourg (Saclay, France: CEA)
\bibitem[Stella \& Vietri(1998b)]{ste-vie:1998:b} Stella, L., \& Vietri, M. 1998b, ApJ, 492, L59
\bibitem[Stella \& Vietri(2001)]{ste-etal:2001} Stella, L. \& Vietri, M. 2001, X-ray Astronomy 2000, Astronomical Society of the Pacific Conference Series,  ed. {Giacconi}, R.,  {Serio}, S. \& {Stella}, L., 213
\bibitem[Stella et al.(1999)]{ste-etal:1999} Stella, L., Vietri, M.,\& Morsink, S.~M. 1999, ApJL, 524, L63
\bibitem[{Stergioulas \& Friedman}(1995)]{ste-fri:1995} Stergioulas, N., Friedman, J. L., 1995,	Astrophysical Journal, Part 1 (ISSN 0004-637X), vol. 444, no. 1, p. 306-311
\bibitem[{Strohmayer} \& {Markwardt}(2002)]{stro-Mar:2002} {Strohmayer}, T.~E., \& {Markwardt}, C.~B. 2002, ApJ, 577, 337
\bibitem[{Stuchl{\'{\i}}k} et al.(2008)]{stu-etal:2008} {Stuchl{\'{\i}}k}, Z., {Konar}, S., {Miller}, J.~C., \& {Hled{\'{\i}}k}, S. 2008, A\&A, 489, 963
\bibitem[{Stuchl{\'{\i}}k} et al.(2013)]{stu-etal:2013} {Stuchl{\'{\i}}k}, Z., {Kotrlov{\'a}}, A., {T{\"o}r{\"o}k}, G. 2013, A\&A,  552, 41
\bibitem[{Stuchl{\'{\i}}k} et al.(2014)]{stu-etal:2014} {Stuchl{\'i}k}, Z., {Kotrlov{\'a}}, A., {T{\"o}r{\"o}k}, G., {Goluchov{\'a}}, K. 2014, AcA, 64, 45
\bibitem[{Stuchl{\'{\i}}k} et al.(2015)]{stu-etal:2015} {Stuchl{\'i}k}, Z., {Urbanec}, M., {Kotrlov{\'a}}, A., {T{\"o}r{\"o}k}, G., {Goluchov{\'a}}, K. 2015, AcA, 65, 169
\bibitem[{Titarchuk} \& {Wood}(2002)]{tit-ken:2002} {Titarchuk}, L., \& {Wood}, K. 2002, ApJ, 577, L23
\bibitem[{T{\"o}r{\"o}k}(2009)]{tor:2009} {T{\"o}r{\"o}k}, G. 2009, A\&A, 497, 661
\bibitem[{T{\"o}r{\"o}k} et al.(2008a)]{tor-etal:2008a} {T{\"o}r{\"o}k}, G., {Abramowicz}, M.~A., {Bakala}, P. et al. 2008a, AcA, 58, 15
\bibitem[{T{\"o}r{\"o}k} et al.(2008b)]{tor-etal:2008b} {T{\"o}r{\"o}k}, G.,  {Abramowicz}, M.~A., {Bakala}, P. et al. 2008b, AcA, 58, 113
\bibitem[{T{\"o}r{\"o}k} et al.(2008c)]{tor-etal:2008c} {T{\"o}r{\"o}k}, G., {Bakala}, P., {Stuchlik}, Z., \& {\v{C}ech}, P. 2008c, AcA, 58, 1
\bibitem[{T{\"o}r{\"o}k} et al.(2010)]{tor-etal:2010} {T{\"o}r{\"o}k}, G., {Bakala}, P., {{\v S}r{\'a}mkov{\'a}}, E., {Stuchl{\'{\i}}k}, Z., \& {Urbanec}, M. 2010, ApJ, 714, 748
\bibitem[{T{\"o}r{\"o}k} et al.(2012)]{tor-etal:2012} {T{\"o}r{\"o}k}, G., {Bakala}, P., {{\v S}r{\'a}mkov{\'a}}, E. et al. 2012, ApJ, 760, 1383
\bibitem[{T{\"o}r{\"o}k} \& {Stuchl{\'{\i}}k}(2005)]{tor-stu:2005} {T{\"o}r{\"o}k}, G., \& {Stuchl{\'{\i}}k}, Z. 2005, A\&A, 437, 775
\bibitem[{T{\"o}r{\"o}k} et al.(2007)]{tor-etal:2007} {T{\"o}r{\"o}k}, G., {Stuchl{\'{\i}}k}, Z., \& {Bakala}, P. 2007, CEJPh, 5, 457
\bibitem[{T{\"o}r{\"o}k} et al.(2014)]{tor-etal:2014} {T{\"o}r{\"o}k}, G., {Urbanec}, M., {Ad{\'a}mek}, K., \& {Urbancov{\'a}}, G. 2014, A\&A, 564, L5
\bibitem[{T{\"o}r{\"o}k} et al.(2016)]{torok-etal:2015:MNRAS:} {T{\"o}r{\"o}k}, G., {Goluchov\'{a}}, K., {Hor\'{a}k}, J. et al. 2016, MNRAS, 457, L19
\bibitem[Urbanec et al.(2010a)]{urb-etal:2010:a} {Urbanec}, M., {B\v{e}t{\'a}k}, E., \& {Stuchl{\'{\i}}k}, Z. 2010a, AcA, 60, 149
\bibitem[{Urbanec} et al.(2013)]{urb-etal:2013} {Urbanec}, M., {Miller}, J.~C., \& {Stuchl{\'{\i}}k}, Z. 2013, MNRAS, 433, 1903
\bibitem[{Urbanec} et al.(2010b)]{urb-etal:2010:b} {Urbanec}, M., {T{\"o}r{\"o}k}, G., {{\v S}r{\'a}mkov{\'a}}, E. et al. 2010b, A\&A, 522, A72 
\bibitem[van der Klis(2005)]{kli:2005} {van der Klis}, M. 2005, AN, 326, 798
\bibitem[{van der Klis}(2006)]{kli:2006} {van der Klis}, M. 2006, Rapid X-ray Variability (UK: Cambridge University Press), 39-112
\bibitem[Wagoner(1999)]{wag:1999} {Wagoner}, R.~V. 1999, PhR, 311, 259
\bibitem[Wagoner et al.(2001)]{wag-etal:2001} {Wagoner}, R.~V.,  {Silbergleit}, A.~S., \& {Ortega-Rodr{\'{\i}}guez}, M. 2001, ApJ, 559, L25
\bibitem[Wang et al.(2015)] {wan-etal:2015:MNRAS:} {Wang}, D.~H., {Chen}, L., {Zhang}, C.~M. et al. 2015, MNRAS, 454, 1231
\bibitem[Watts(2012)]{wat:2012:ARAA} Watts, A.~L. 2012, \araa, 50, 609
\bibitem[{Wiringa} et al.(1988)]{wir-etal:1988} {Wiringa}, R.~B., {Fiks}, V., \& {Fabrocini}, A. 1988, PhRvC, 38, 1010
\bibitem[{Zhang}(2005)]{zha:2005} {Zhang}, C.-M. 2005, ChJAS, 5, 21
\end{thebibliography}
\end{document}